\documentstyle[preprint,aps,epsf]{revtex}

\def\be{\begin{equation}}
\def\ee{\end{equation}}

\def\o{\over}
\def\ba{\begin{array}}
\def\ea{\end{array}}
\def\bea{\begin{eqnarray}}
\def\eea{\end{eqnarray}}

\begin{document}
\title{Can radiative correction cause large neutrino mixing?}
\author{Gautam Dutta\footnote{email: gautam@mri.ernet.in}}
\address{Harish Chandra Research Institute, Jhusi, Allahabad 211019, India}
\maketitle
\begin{abstract}
We investigate whether radiative corrections can be
responsible for the generation of large mixing in a pair of degenerate
neutrino with same CP parity. We find that this mechanism is fine tuned and
doesn't work for arbitrary mixing at the high scale. 
\end{abstract}

The data from the solar neutrino and the  atmospheric neutrino experiments 
can be explained through flavor oscillations of massive
neutrinos. The favoured solution of both these 
problems are a pair of neutrinos with very small mass squared difference and
large mixing\cite{fukuda,bahcall}. 
The mass square difference required for the solar neutrino
oscillation is less than $10^{-5}eV^2$ while for the atmospheric neutrinos
it is $10^{-3}eV^2$. 
If neutrinos form the hot Dark Matter then the masses of all
the neutrinos have to be of the order of 1eV. This, along with the very small
mass square differences suggests that neutrino
mass spectrum is almost degenerate. 
If neutrinos have Majorana
mass, then one can have a pair of degenerate neutrinos with opposite CP
parities. These are called Pseudo Dirac neutrinos. The mixing between such
pairs can be shown to be maximal \cite{wolfenstein}. 
The smallness of neutrino mass can be
obtained through the seesaw mechanism\cite{seesaw} in which neutrinos can have both Dirac
and Majorana masses. Even in this mechanism Pseudo Dirac neutrinos with
maximal mixing are possible for a suitable choice of the structure of right
handed Majorana mass matrix\cite{dutta}. 
While these models can generate only one maximal
mixing in 3 generation scenario, the solar neutrino problem and the
atmospheric neutrino problem requires two pairs with maximal
mixing. 
Another problem is that for degenerate Majorana neutrinos, large mixing may
not be stable under small radiative corrections. A pair of Majorana
neutrinos with opposite CP parities is stable but for a pair with same CP
parity large mixing collapse to 0 due to radiative correction\cite{ma}. 
A detailed analysis on the stability of the neutrino mass matrix and mixing
for various type of mass hierarchy is done in \cite{haba,chankowski,ibarra}.   
Some authors \cite{balaji} 
suggested that the same instability can cause large
mixing at the low scale from an arbitrary mixing at the high scale. This
mechanism is similar to the MSW mechanism where the mixing angle gets
enhanced at the resonance. As this requires the present scale of the universe
to be at the `resonance' as far as neutrino mixing is concerned, this
is a fine tuned solution. This is because 0 mixing is a fixed point of the
renormalisation group evolution of neutrino masses and mixing whereas
maximal mixing is not \cite{chankowski}. 
The fact that for degenerate mass structure, large mixing is unstable have
been discussed by other authors \cite{haba,chankowski,ibarra,ellis}
We investigate whether such a
mechanism can be at work to generate the required large mixing considering
the range of mass square differences from the solar and atmospheric neutrino
experiments.  

Let $m_1$ and $m_2$ be the mass eigenvalues of two neutrinos at a high scale
$\Lambda$. Let $\theta$ be the mixing angle at this scale. Then in the
flavour basis the mass matrix will be given as
\[
M_F(\Lambda)=\left(\ba{rr}
   m_1\cos^2\theta + m_2\sin^2\theta & {m_2-m_1 \o 2}\sin(2\theta)  \\
    {m_2-m_1 \o 2}\sin(2\theta) &  m_1\sin^2\theta + m_2\cos^2\theta \ea 
                                                   \right)
                                                          \nonumber
\]
Due to radiative correction this mass matrix gets modified in the low
scale $M_Z$ to \cite{ellis,babu,chankowski1}
\bea
M_F(M_Z)&=&\left(\ba{cc} 1+\delta_\alpha & 0 \\ 0 & 1+\delta_\beta \ea \right ) 
         M_F(\Lambda)
     \left(\ba{cc} 1+\delta_\alpha & 0 \\ 0 & 1+\delta_\beta \ea \right )
                                                 \nonumber       \\
 &=& \left(\ba{rr} (1+2\delta_\alpha)( m_1\cos^2\theta + m_2\sin^2\theta ) &
    (1+\delta_\alpha + \delta_\beta){m_2-m_1 \o 2}\sin(2\theta) \\
    (1+\delta_\alpha + \delta_\beta){m_2-m_1 \o 2}\sin(2\theta)  &
    (1+2\delta_\beta)( m_1\sin^2\theta + m_2\cos^2\theta ) \ea \right)
						\nonumber   \\
 &\equiv & \left(\ba{cc} A & B \\ B & C \ea \right)     \label{abc}
\eea

Here $\delta_\alpha$ and $\delta_\beta$ are the radiative corrections to 
the flavours $\alpha$ and $\beta$.

The mixing angle at the low scale $\theta'$ in the limit of near degeneracy
$m_1\approx m_2 \approx m$ is given as 
\bea
\tan 2\theta'&=&{2B\o C-A} \label{tan} \nonumber \\
  & =& 
  {(1+\delta_\alpha+\delta_\beta)(m_2-m_1)\sin (2\theta) \o
  (m_2-m_1)\cos(2\theta) + 2(\delta_\beta-\delta_\alpha)m} \label{theta'}
\eea
If 
\be
2(\delta_\beta-\delta_\alpha) = {m_1-m_2 \o m}\cos(2\theta) \label{cond}
\ee
 then
$\theta'\to 45^o$

Let $\epsilon=2(\delta_\beta-\delta_\alpha)$. Without loss of generality let
$m_1 > m_2$. 
Whatever mixing angle $\theta$ one takes 
at high scale
$\Lambda$, when one goes to low scale the mixing angle $\theta'$ will be
maximal if condition (\ref{cond}) is satisfied\cite{balaji}. We call this
value of $\epsilon$ given by eq.(\ref{cond}) as $\epsilon_r$ ($r$ referring
to resonance). 
However it is crucial to investigate whether this condition can be satisfied
at an arbitrary mixing angle $\theta$ at the high scale.  
This is because the masses $m_1$ and $m_2$ in eq.(\ref{cond})
are the masses at the high scale $\Lambda$. Though the masses
themselves don't have significant evolution through radiative correction the
same cannot be said about the mass differences near degeneracy. 
To see this, consider the mass square difference at the low scale
\[
\Delta m^2(M_Z) = (C+A)\sqrt{(C-A)^2 + 4B^2}   
\]
From eq.(\ref{tan}) this becomes 
\[
\Delta m^2(M_Z)\sin(2\theta') = 2B(C+A)         
\]
From eq.(\ref{abc}), in the limit of near degeneracy we get
\be  
\Delta m^2(M_Z)\sin(2\theta')= \Delta m^2(\Lambda)\sin(2\theta)
                                                 \label{dm2mz}
\ee
Here we have used the near degeneracy approximation in the following
\[
{m_2-m_1 \o m}={(m_2 - m_1)(m_2 + m_1)\o m(m_2 + m_1)}\approx
   {\Delta m^2 \o 2m^2}                            
\]
This is as expected
in a MSW like scenario, where due to significant change in mass square
difference, the mixing angle undergoes drastic enhancement in the resonance
region. Eq.(\ref{dm2mz}) says that once $\Delta m^2(M_Z)$ and $\sin(2\theta')$
are determined from oscillation experiments, the mass square difference and
the mixing angle at the high scale $\Lambda$ are related. 
So if the mixing
angle $\theta$ is small at the high scale then the mass square difference
$\Delta m^2(\Lambda)$ should be large. This would need a large $\epsilon$  to
satisfy the resonance condition (\ref{cond}). We will see this in detail in
the numerical plots below.    

When $\theta'$ is  maximal, we have from (\ref{dm2mz}) 
\be 
\Delta m^2(M_Z)=\Delta m^2(\Lambda)\sin(2\theta)   \label{dm2mzmax} 
\ee
We see here that once the evolution of $\Delta m^2$ is fixed from the scale
$\Lambda$ to the scale $M_Z$, the mixing angle at the high scale $\Lambda$ is
fixed and not arbitrary.

Eq.(\ref{cond}) is a relation between $\epsilon_r$
and the mass difference and mixing angles at the high scale. It would be
appropriate to get $\epsilon$ and $\epsilon_r$ in terms of the 
quantities at the low scale.
Rearranging eq.(\ref{tan}) and using eq.(\ref{dm2mz}) we get  
\be
\epsilon={\Delta m^2(M_Z) \o 2m^2} \sin(2\theta')
         \left({1 \o \tan(2\theta')} - {1 \o \tan(2\theta)}\right)  
                                                              \label{epst'}
\ee
Since the combination $\Delta m^2(M_Z)\sin(2\theta')$ does not change under 
renormalisation group evolution as evident from eq.(\ref{dm2mz}), we see
from eq.(\ref{epst'}) that given an $\epsilon$ if a small $\theta$ is enhanced
to a large $\theta'$ around $45^o$ then a large $\theta$ will end up at a 
$\theta'$ around $90^o$ which is small mixing but with inverted hierarchy.
This observation is consistent with the explicit evolution of the neutrino
mixing in SM, MSSM and two Higgs doublet model done in ref. \cite{babu} 
and \cite{haba1}.

$\epsilon_r$ is given by $\theta'=45^o$ in eq.(\ref{epst'})
\be
\epsilon_r={\Delta m^2(M_Z) \o 2m^2} \left( - {1 \o
                                          \tan(2\theta)}\right) \label{cond1}
\ee
From eq.(\ref{epst'}) and (\ref{cond1}) we get the range of $\epsilon$ that
generates large mixing around the maximal. This is given as
\be
\delta\epsilon =\epsilon_r - \epsilon 
                \approx {\Delta m^2(M_Z) \o 2m^2} 
                        \left( {\sin(2\theta') - 1 \o
                        \tan(2\theta)}\right)            \label{epsrange}
\ee
where $\tan(2\theta') >> \tan(2\theta)$ around $\theta'=\pi/4$ and small
$\theta$. So if we want the
mixing at the low scale to be anything in the range of $30^o$ to $45^o$ the
range of $\epsilon$ can be obtained by putting $\theta'=30^o$ in 
eq.(\ref{epsrange}).
This will be about $15\%$ of $\epsilon_r$. 
If we also consider the reversed hierarchy
after maximal mixing, this range becomes $30\%$. 
With the current possibility of atmospheric neutrino mixing angle
to be very near maximal $(40-50)^o$, this range gets restricted to about 
$4\%$ around $\epsilon_r$.

\begin{figure}[p]
%\special{psfile="range.ps"
%         hscale=100 vscale=100 hoffset=00 voffset=-200}
\centerline{
\epsfxsize=10cm \epsfysize=15cm
\epsfbox{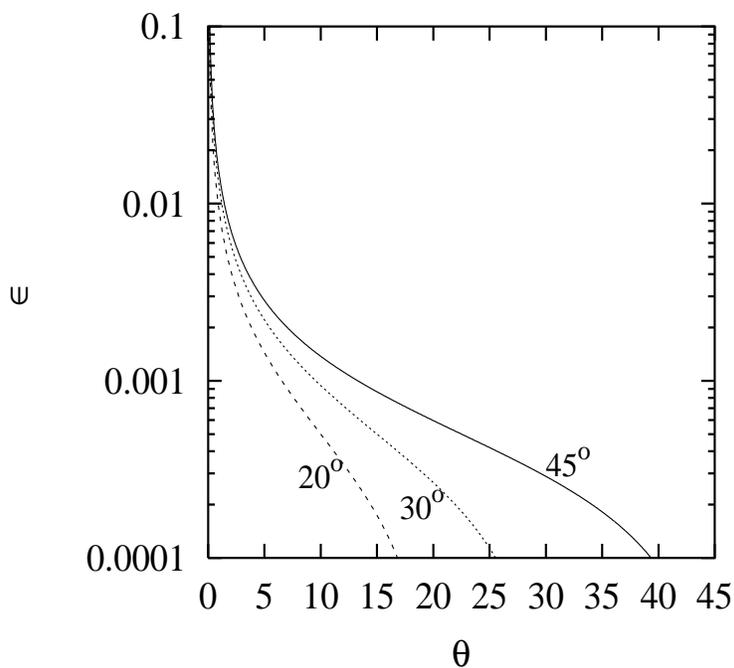}
           }
\caption{$\epsilon$ required to generate $\theta'$ at the low scale as a
function of $\theta$ at the high scale. The solid line, dotted line and the
dashed lines are  for $\theta'=45^o, 30^o$ and $20^o$ respectively. Here
$\Delta m^2(M_Z) = 10^{-3}eV^2$.}

\label{fig1}
%\vskip 18cm
\end{figure}

While it may be possible for this mechanism to enhance some mixing angle 
from high scale to low scale with a few $\%$ range of $\epsilon$ around
$\epsilon_r$, another question
we ask is, given an $\epsilon$ what is the range of mixing angle at the
high scale that can be enhanced at the low scale. 
Figure (\ref{fig1}) shows a plot of $\epsilon$ verses the high scale 
mixing angle
$\theta$ at various values of the low scale mixing $\theta'$ as obtained
from eq.(\ref{epst'}). We have taken $\Delta m^2(M_Z) = 10^{-3}eV^2$ and
$m=1eV$. So this is the case for atmospheric neutrino solution.
It is clear from the figure that for very small $\theta$ ($<0.5^o$) the
$\epsilon$ needed for the enhancement to occur is as large as 0.1. 
As $\epsilon$ is
expected to be small, this mechanism cannot produce a large mixing from
such small mixing at high scales. If $\epsilon \sim 0.01$ only $\theta$ in
a small range $1.5^o-2^o$ can be enhanced to near maximal mixing ($30^o$ to
$45^o$) at the low scale. But if $\theta$ is over $2.5^o$ then this
mechanism becomes an overkill as the low scale mixing then again becomes
small but with an inverted hierarchy. This is as expected in MSW mechanism
when the neutrinos pass through the resonance and then go far away from it,
the mixing is no longer maximal and the hierarchy gets inverted. If
$\epsilon \sim 0.001$, then only $\theta$ between $10^o$ to $15^o$ get 
enhanced.
While for $\epsilon=0.0001$ high scale mixing upto about $25^o$ don't get
any significant enhancement. If the same analysis is done for the solution
of solar neutrinos then the corresponding $\epsilon$ should be lowered by
two orders of magnitude as $\Delta m^2(M_Z)$ is $\sim 10^{-5}eV^2$ instead of
$10^{-3}eV^2$ in eq.(\ref{cond1}). So we see that there is no range of
$\epsilon$ that can enhance a large range of mixing angles at the high
scale. Thus in order to generate maximal mixing at the low scale through
radiative correction of the masses at the high scale, one has to start with
specific mixing angle at the high scale depending upon the particular model
one considers. Similar conclusion have been obtained in specific models
\cite{ibarra,babu,haba1} through
renormalisation group evolution of neutrino mass matrix.

To summarise, we see that radiative correction cannot give a general
enhancement of any mixing angles at the high scales to maximal mixing at the
low scale. If it works due to fine tuning of parameters within a few percent 
for low mixing
angles, it would kill the large mixing at high scales to small
mixing at low scale. This is somewhat to be expected as large mixing between
neutrinos of same CP parity is unstable near mass degeneracy
\cite{ma,haba,chankowski,ibarra}. 
Radiative magnification doesn't give the freedom to start with arbitrary
mixing angle at high scale in building models of neutrino mass and mixing
structures. 

%\begin{figure}[htp]

\end{document}